\documentstyle[epsfig]{aipproc}

\def\lta{\;\rlap{\lower 2.5pt                       
             \hbox{$\sim$}}\raise 1.5pt\hbox{$<$}\;}

\def\msun{M_{\odot}}

\begin{document}
\title{X-Ray Observations of Low-Mass X-Ray Binaries:
Accretion Instabilities on Long and Short Time-Scales}

\author{Jean H. Swank}
\address{Laboratory for High Energy Astrophysics\\NASA/GSFC Greenbelt, MD 
20771
}

\maketitle

\begin{abstract}
X-rays trace accretion onto compact objects in binaries with low mass
companions at rates ranging up to near Eddington.  Accretion at high
rates onto neutron stars goes through cycles with time-scales of days
to months. At lower average rates the sources are recurrent
transients; after months to years of quiescence, during a few weeks
some part of a disk dumps onto the neutron star.  Quasiperiodic
oscillations near 1 kHz in the persistent X-ray flux attest to
circular motion close to the surface of the neutron star. The neutron
stars are probably inside their innermost stable circular orbits and
the x-ray oscillations reflect the structure of that region. The long
term variations show us the phenomena for a range of accretion rates.
For black hole compact objects in the binary, the disk flow tends to
be in the transient regime. Again, at high rates of flow from the disk
to the black hole there are quasiperiodic oscillations in the
frequency range expected for the innermost part of an accretion
disk. There are differences between the neutron star and black hole
systems, such as two oscillation frequencies versus one. For both
types of compact object there are strong oscillations below 100 Hz.
Interpretations differ on the role of the nature of the compact
object.

\end{abstract}

\section*{Introduction}

Low-mass X-ray binaries (LMXB) are the binaries of a low-mass
``normal'' star and a compact star. The compact star could be a white
dwarf, a neutron star, or a black hole.  The Rossi X-Ray Timing
Explorer ({\it RXTE}) has been observing since the beginning of 1996
and has obtained qualitatively new information about the neutron star
and black hole systems. In this paper I review the new results briefly 
in the
context of what we know about these sources. The brightest, Sco X--1,
was one of the first non-solar X-ray sources detected, but only with
{\it RXTE} have sensitive measurements with high time resolution been
made that could detect dynamical time-scales in the region of strong
gravity. {\it RXTE} also has a sky monitor with a time-scale of hours
that keeps track of the long term instabilities and enables in
depth observations targeted to particular states of the sources.

The LMXB have a galactic bulge or Galactic Population II
distribution. The mass donor generally fills its Roche lobe, is less
than a solar mass, and is optically faint, in contrast to the early
type companions of pulsars like Cen X--3 or the black hole candidate
Cyg X--1. In many cases the optical emission is dominated by emission
from the accretion disk, and that is dominated by reprocessing of the
X-ray flux from the compact object \cite{vPM95}.  
The known
orbital periods of these binaries range from 16 days (Cir X--1) to 11
minutes (4U 1820--30). The very short period systems ($< 1$ hr) are
expected to have degenerate dwarf mass donors and probably the mass
transfer is being driven by gravitational radiation. The different
properties of the sources indicate several populations. The longer
period systems with more massive companions are probably slightly
evolved from the main sequence.

There are about 50 persistent neutron star LMXB \cite{vP95}.
Distances can be estimated in a variety of ways. The hydrogen column
density indicated by the X-ray spectrum should include a minimum
amount due to the interstellar medium. Many of the sources emit X-ray
bursts associated with thermonuclear flashes that reach the hydrogen or
helium Eddington limits. In some cases the optical source provides
clues. The resulting luminosity distribution appears to range from
several times the Eddington limit for a neutron star down below the
luminosity of about $10^{35}$ ergs s$^{-1}$, corresponding to $\approx 
10^{-11}$
$\msun$ yr$^{-1}$ \cite{CS97}.  The lower limit has come from instrument
sensitivity, but it may also reflect the luminosity below which the
accretion flow is not steady, so that the source must be a transient.

``X-Ray Novae'' that are among the brightest X-ray sources for a month
to a year are sufficiently frequent that they were seen in rocket
flights in the beginning of X-ray astronomy. The X-ray missions that
monitored parts of the sky during the last three decades found that on
average there are 1--2 very bright transient sources each year
(e.g. \cite{CSL97}) with durations of a month to a year.  
In 5 years of {\it RXTE} operations,  
we know of 20 transient neutron
star sources and and an equal number of transient black hole
sources. If they have a 20 yr recurrence time we have seen only a quarter
of them and if we have only been watching a third of the region in the
sky, 20 observed sources implies more than 240 sources exist. In
reality there is a distribution of the recurrence times, some as short
as months, others longer than 50 years, if optical records are good.
On the basis of such estimates, the number of potential black hole
transients is estimated to be on the order of thousands \cite{TS96}.

The separation of sources into persistent and transient sources is a
very gross simplification. One of the discoveries of recent missions,
and especially of the All Sky Monitor (ASM) \cite{Bradt00} has been that the
persistent sources have cycles of variations with time-scales ranging
from many months to days.  If the transient outbursts originate in
accretion instabilities, perhaps these variations are related. 
In the next section I show some of the kinds of behavior being observed. 

At radii close to the compact objects the dynamical time-scale gets
shorter, till it is the milliseconds of the neutron star or black
hole. RXTE's large area detectors detect oscillations on these
time-scales which must reflect the dynamics at the innermost stable
circular orbit (ISCO) of these neutron stars and black holes.

The neutron stars of this sample are expected to have magnetic dipole
moments and surface fields about $10^8 - 10^9$ gauss. Of course the
neutron stars have a surface such that matter falling from the
accretion disk to the neutron star crashes into the surface and
generates X-ray emission. In the case of the black holes matter could
fall through the event horizon and disappear with no further emission
of energy.  Thus the X-rays produced and the dynamics that dominates
in the two cases (neutron star versus black hole) could be
different. However, a number of similarities appear in the signals we
receive.

\section*{Long Time-scale Variabilities}
 
\subsection*{High Accretion Rate - Persistent Sources}

Among the persistent LMXB there are characteristic variations on
time-scales of months in some sources and days in
others\cite{Bradt00}. Quasiperiodic modulations were pointed out at 
37 days for Sco~X--1 (IAUC 6524), 
24.7 days for GX 13+1 (IAUC 6508), 
77.1 days for Cyg~X--2 (IAUC 6452), 
37 days for X 2127+119 in M15 (IAUC 6632). The
obviously important, but not strictly periodic modulations in 4U
1820--30 and 4U 1705--44 at time-scales of 100--200 days are shown in
Figure 1. For Sco X--1, the changes in activity level occur in a day and
the activity time-scale is hours. The hardness is often correlated
with the rate, although this measure does not bring out more subtle 
spectral changes.

These time-scales are less regular than the 34 day cycle time of Her
X-1, and similar modulations in LMC X-4 and SMC X-1, which are thought
to be due to the precession of a tilted accretion disk.  The latter
sources are high magnetic field pulsars in which the disk is larger
than in the LMXB, and is truncated by the magnetosphere at a
radius as large as $10^8$ cm. The LMXB spectral changes are also
different than those of the pulsars. In the LMXB case the changes are
thought to be real changes in the accretion onto the
neutron star, at least the production of X-rays, rather than a change
in an obscuration of the X-rays that we see.

The spectral changes are captured in the color-color diagrams that
give rise to the names ``Z'' and ``Atoll'' for subsets of the LMXB.
These were identified with EXOSAT observations by Hasinger and van der Klis
\cite{HvdK89}.  Characteristics of the bursts from 4U 1636--53 depended on the
place of the persistent flux in the atoll color-color diagram \cite{vdK90}. 
This
implied that the  real mass accretion rate was
correlated with the position on the diagram (although other possibilities
such as the distribution of accreted material on the surface of the
neutron star may play a role). That the position in the diagram
in not uniquely correlated to the flux is as yet not understood.
Transients atoll sources like Aql X--1 and 4U 1608--52
go around the atoll diagram during the
progress of the outburst.

\begin{figure}

\centerline{\epsfig{file=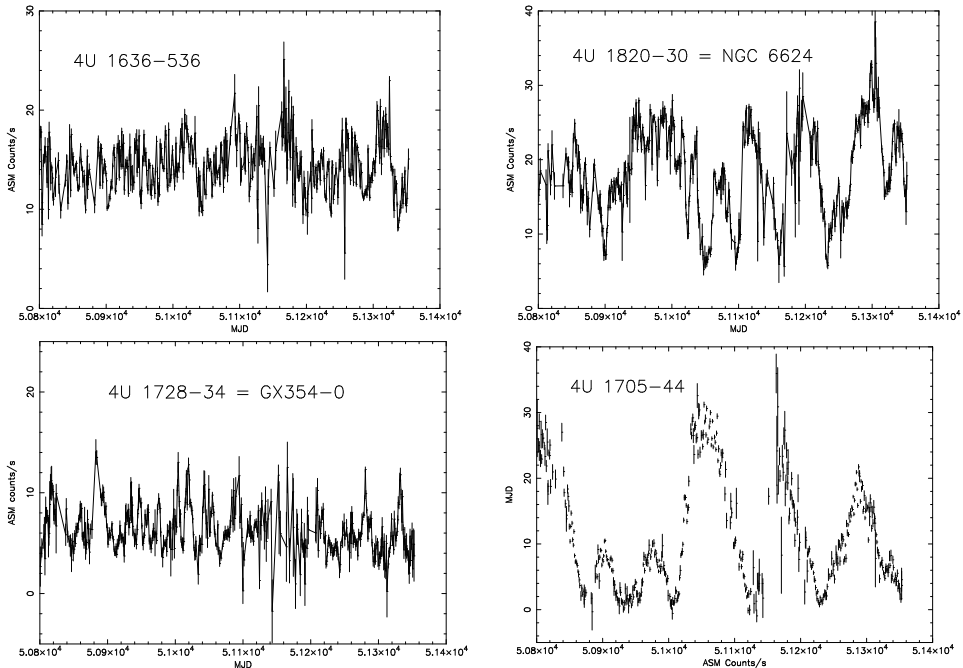,angle=-90,width =5.0in}}
\caption{{\it RXTE} ASM Rates from Four Atoll Bursters. 
Modulations  are 
typically a factor of two, although sometimes more. Many properties 
vary with these modulations.}
 \label{fig1}

\end{figure}

\subsection*{Low Average Accretion Rate - Transients}

There are only a few persistent LMXB in which the compact object is a
black hole. Black hole binaries are for some reason more likely to be
transients. Perhaps the binaries harboring them are not being driven
to have as much mass exchange, so that it happens that these systems
are in the range of mass flow through the disk that makes them
transient. There are also neutron star transients with low average
mass exchange rates. Figure 2 shows on the left two neutron star
transients, a well known atoll burster Aql X--1 and the pulsar GRO
J1744--28, which had two outbursts a year apart, but has otherwise not
been seen. On the right are two black hole candidates, 4U 1630--47,
which recurs approximately every two years, and XTE J1550--564, which
like GRO J1744--28, had a dramatic outburst, with a weaker recurrence
after a year's hiatus. Black hole candidates can get brighter than the transient bursters,
consistent with the Eddington limit for more massive compact objects
and they probably go through more different spectral and timing
``states'', but there are also similarities in the kinds of behavior
that are exhibited.

\begin{figure}[t]

\centerline{\epsfig{file=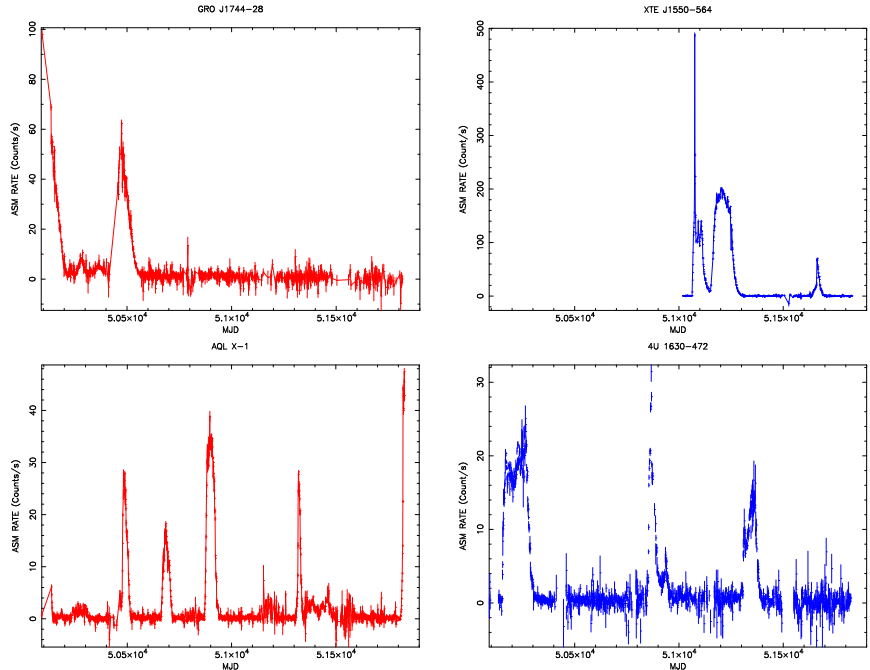,angle=-90,width=4.5in}}
\caption{{\it RXTE} ASM Rates for Four Transients. 
(left) Neutron Star Transients GRO 1744-28 (top) and Aql X-1 (bottom). 
GRO J1744-28 was bright at the launch of {\it RXTE}. 
(right) Black hole candidates XTE J1550-564 (top) and 4U 1630-472 (bottom). 
Note that the figures have different scales for the ASM counts s$^{-1}$. 
(1 Crab = 75 ASM counts s$^{-1}$.)}
\label{fig2}

\end{figure}

From both BeppoSAX and RXTE results it is clear that there is a
population of systems which have transient episodes, but which
are an order of magnitude less luminous at peak. BeppoSax has seen
bursts from a number of sources for which the persistent flux is below
their sensitivity limit. RXTE has seen a dozen sources which may not
rise above $10^{36}$ ergs s$^{-1}$ during transient episodes. Several
of these are believed to be neutron stars because Type I (cooling)
bursts were observed.  They include the source SAX J1808.4-3658,
unique to date, that both pulses (2.5 msec) and has Type I bursts.

Some sources have spectral and timing properties consistent with black
hole candidates which go into the black hole ``low hard'' state, with
strong white noise variability below 10 Hz and hard spectra. One of
these was V4641 Sgr, which went into much brighter outburst, with a
radio jet, before disappearing.

\section*{Instabilities Close to the Compact Object}

\subsection*{Kilohertz  Oscillations for Neutron Stars - near the ISCO}

More than 22 LMXB have now exhibited a signal at kilohertz frequencies
in the power spectra of the x-ray flux (See \cite{vdK00}).  Figure 3
(thanks to T. Strohmayer) shows results for samples of data from an atoll
and a Z source.  Usually this signal is two peaks at 1--15 \%
power. They indicate quasi-periodic oscillations with coherence (mean
frequency/frequency width) as much as 100.  The centroid frequencies
are not constant for a source, but vary. Over a few hours the
frequency is correlated with the X-ray flux, increasing with the
flux. The flux variations of a factor of two are correlated with
changes of frequency between 500 Hz and 1000 Hz, approximately
\cite{SSZ98}. The highest reported is 1330 Hz, from 4U 0614+09.
Considering that for a circular orbit at the Kepler radius $r_K$, the
observed frequency is $(2183/M_1) (r_{ISCO}/r_K)^{3/2}$, where
$r_{ISCO} = 6GM/c^2$ is the innermost stable circular orbit for a
spherical mass $M = M_1 \msun$ of smaller radius, neutron stars of
masses $M_1$ = 1.6--2.0 would have Kepler frequencies at the ISCO of
just such maximum frequencies as are observed.

\begin{figure}[t]

\centerline{\epsfig{file=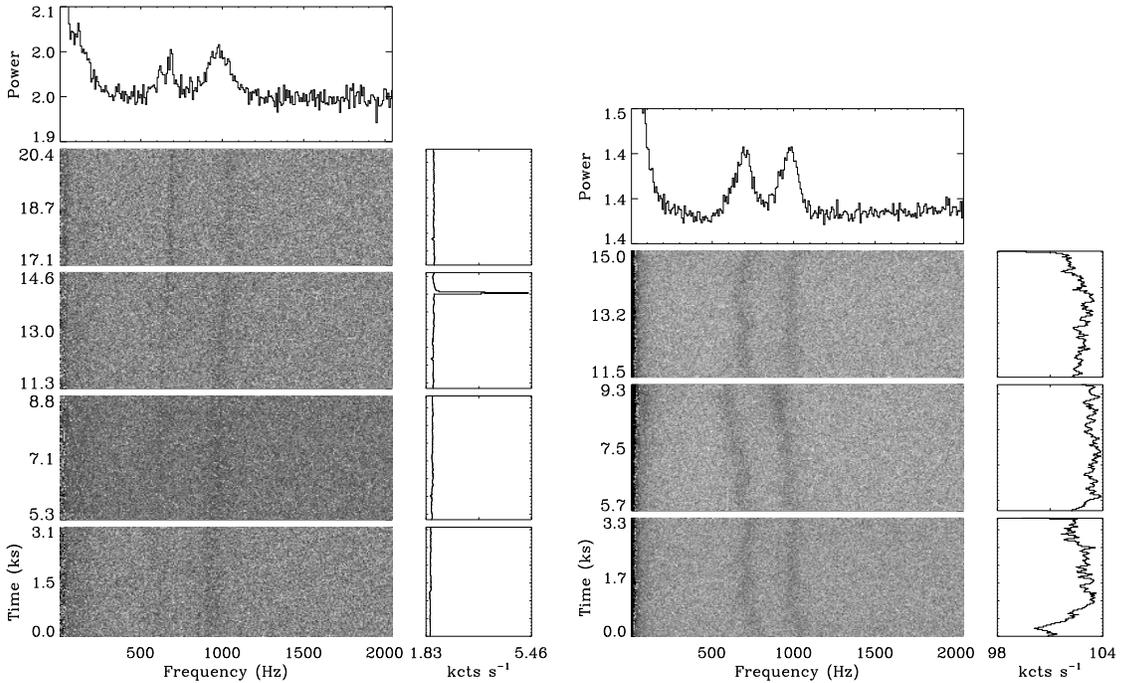,width=6.0in}}
\caption{{\it RXTE} Power Density Spectra. (left) Atoll Source
4U1728-34.  (right) Z Source Sco X-1. For each the grey scale plot of
power as a function of frequency and time is shown for sequential
observing intervals. The average PDS is shown above and the count rate
during the observations on the right. One burst occurred during the 4U
1728-34 observation.  The power spectra used are for 32 s data
intervals.  }
\label{fig3}

\end{figure}

While the luminosities of the sources exhibiting these QPO range from
$10^{36}$ ergs s$^{-1}$ to above $10^{38}$ ergs s$^{-1}$, the maximum
values of the upper frequency range only between 820 Hz and 1330
Hz. This suggests \cite{Zhang98,Kaaret99} that it
represents a characteristic of the neutron stars fairly independently
of the accretion rate. The ISCO and the neutron star radius are
candidates. For lower fluxes, the frequencies, at least locally in the
light curve, decline, as if the Keplex orbit were further out. Which
is more likely, that the inner radius is then at the ISCO or at the
radius of the neutron star? In the latter case the neutron star is
outside the innermost stable circular orbit.  Understanding the
boundary requires consideration of the radiation pressure, the
magnetic fields, and the optical depth of the inner disk. For sources
with flux near the Eddington limit, the optical depth of the material
near the surface should be much larger than the optical depth of the
material accreting at rates 100 times less. For the inner disk being
at the ISCO, and fairly compact neutron stars, this plausibly does not
matter. For the inner disk at the surface or a large neutron star, it
seems hard to explain the similarity of appearance between luminous Z
sources and fainter atoll sources. There are in fact
differences in the appearance of the QPOs; one is that the amplitude
of the QPOs is larger for the atoll sources than for the Z sources. So
the situation is not completely clear. 

If a disk is truncated at an inner radius which moves in toward the
neutron star as the mass flow through the disk increases and a QPO is
generated at near this inner edge, the frequency would be likely to
increase with the luminosity. The frequency would not be able to
increase beyond the value corresponding to the minimum orbit in which
the disk could persist. Miller, Psaltis, and Lamb \cite{MLP98}
argued that if
radiation drag was responsible for the termination of the disk,
optical depth effects would lead to the sonic point radius moving in
as the accretion rate increases.  
There would be a highest frequency corresponding to the
minimum possible sonic point radius. In the cycles of 4U 1820-30 the
frequency approached a maximum which it maintained as the flux
increased further before the feature became too broad to detect. This
kind of behavior would arise from a sonic point explanation.

From Figure 4, it can be seen that if the equation of state (EOS) of
the nuclear matter at the center of a neutron star is very stiff, near
the L equation of state, for $1.4-2 \msun$ neutron stars the radius of
the star is close to its own ISCO; whether it is inside or outside it
is depends sensitively on the mass. If the equation of state is
softer, closer to the FPS EOS, interpretation of the maximum
frequencies observed as a Kepler frequency {\em at the surface} would
imply a mass significantly less than the $1.4 \msun$ with which many
neutron stars are probably formed. In either case, moderately stiff
EOS and maximum frequency at the ISCO, or stiff equation of state and
maximum frequency either at the ISCO or the surface, the frequency
would be from near the ISCO, if not just outside it. Accurate
considerations require the rotation rate of the neutron star to be
taken into account.

\begin{figure}[t]

\centerline{\epsfig{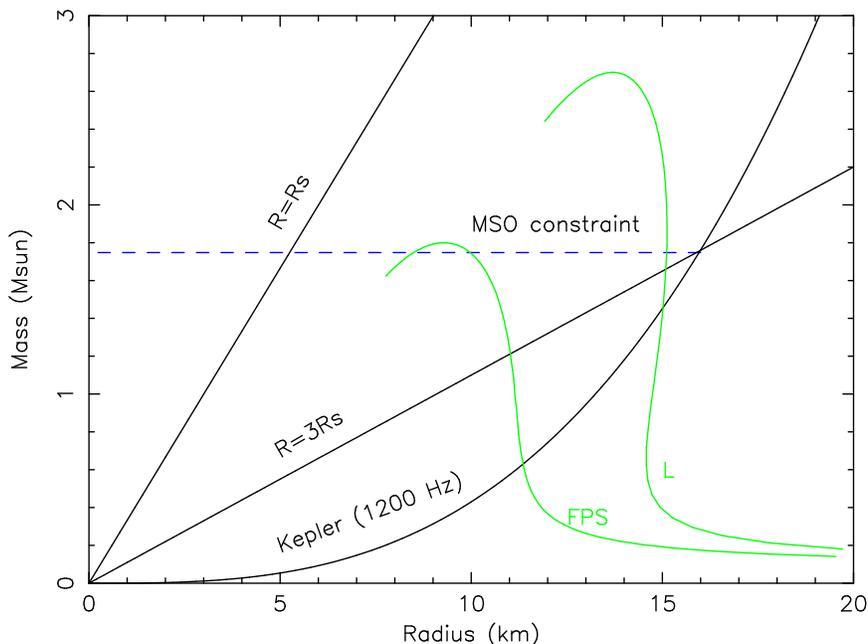}}
\caption{Constraints on the Neutron Star Equation of State 
from the kilohertz QPO. If the frequency of 1200 Hz is a Kepler frequency,
the Mass of the star and the radius of the Kepler orbit are constrained. The orbit cannot be inside the ISCO. If it is at the ISCO the mass is determined and 
any EOS inside that radius which allow a big enough mass would support it. }
 \label{fig4}

\end{figure}

A characteristic of the twin kilohertz peaks is that when the
frequency changes, the two frequencies approximately move together,
with the difference approximately constant, at least until near the
maximum frequencies (and luminosities) for which they are observed in
a given source. This suggests a beat frequency and the relation
between the difference frequency and the frequencies seen during
bursts (See Strohmayer, this volume) suggest the neutron star spin as
the origin of the beats. Miller, Lamb and Psaltis \cite{MLP98} explored 
how the two frequencies could be generated and Lamb and Miller 
refined the model in agreement with the 5 \% changes in the frequency
separation, that are observed \cite{LMiller00}. However, this varying
separation between the two QPO also suggested identification
as the radial epicyclic frequency of a particle moving in an
eccentric orbit in the field of the neutron star. 
The lower of the two frequencies is then identified, not with a beat
frequency, but with the precession of the periastron \cite{Stella99}, 
although efforts to fit the predictions of this
model in terms of particle dynamics produce implausibly large
eccentricities, neutron star masses and spins \cite{MarkovicL00}. Psaltis 
and Norman proposed that similar
frequencies could be resonant in a
hydrodynamic disk \cite{PN99}. In these models, at least
in their current forms, the difference between the two 
QPO peaks is not related
to the spin, but to something like the radial epicycle
frequency.

A quite different class of models are those in which the disk has a
boundary layer with the neutron star and the plasma is excited by
the magnetic field of the neutron star \cite{TOK99}. The magnetic pole 
makes a small angle with the
neutron star rotation axis. In this case
the lower kilohertz QPO frequency is the Kepler frequency, while both
the upper frequency and the low frequency oscillation (corresponding to
the Horizontal Branch Oscillations in Z sourses) are related to
oscillations of plasma interacting with the rotating magnetic field.

\subsection*{Hectohertz oscillations for Black Holes}

Although accreting neutron stars and black holes should have important
differences, they both presumably have an accretion disk with an inner
radius, when the mass flow is high enough. Possible signals from the
ISCO of black holes were discussed when accretion onto black holes 
was first considered \cite{Suny73} and anticipation 
of {\it RXTE} inspired detailed 
calculations \cite{NW93}. 
The {\it RXTE} PCA has detected QPO in 5 black hole candidates at
frequencies that are suitable to be signals from the ISCO of black
holes in the range of $5 - 30 \msun$. They have been observed only in
selected observations and are generally of lower amplitude (a few \%)
than the neutron star kilohertz QPO. For GRS 1915+105, the frequency
has always been 67 Hz \cite{Morgan97} . For GRO 1655--40, Remillard 
identified 300 Hz \cite{1655R99}. 
For XTE~J1550--564, at different times it has been between 185 and 205
Hz \cite{1550R99}. For XTE J1859--262, a broad signal at 200 Hz is
observed in the bright phases near the peak of the outburst
\cite{Cui00}.  For 4U 1630--47 as well, which has had 3 outbursts
during the {\it RXTE} era, Remillard has reported 185 Hz. The black hole
candidates have appeared to differ from the neutron stars in having
one QPO rather than two. An obvious question is whether the second QPO 
is associated with the  presence of a neutron star with a surface 
and a rotating magnetic dipole. Recent work by Strohmayer \cite{Stroh01}
casts doubt on it. 

There were other black hole candidates observed with {\it RXTE}, which did
not exhibit high frequency oscillations and the properties
of the high frequency signal are not very well defined. Interpretation
in terms of Kepler frequency at the ISCO, non-radial g-mode
oscillations in the relativistic region of the accretion disk, and
Lense-Thirring precession have been discussed.  GRO~J1655-40 
is very interesting because the radial velocities of
absorption lines of the secondary have given rather precise
measurement of the mass. (The best estimates are so far $5.5-7.9
\msun$ \cite{Shahbaz99}.) In this case the mass  well
known and the black hole's angular momentum can be the goal. The 300 Hz
frequency is high enough that for a g-mode the black hole would have
near maximal angular momentum, but if it represents a Kepler velocity, a 
Schwarzschild black hole would still be possible\cite{Wagoner98}. 
The question has been
asked whether the microquasars GRS 1915+105 and GRO J1655-40 have
powerful radio jets associated with outbursts because they have fast
rotation \cite{Mirabel99}.

\subsection*{Decahertz Oscillations for Neutron Stars and Black Holes}

In the Z source LMXBs the first QPOs discovered were the
Horizontal Branch Oscillations (HBO), first seen by EXOSAT, but then by
Ginga. They occur in the range 15--50 Hz, have amplitudes as high as 30 \%,
increase in frequency with the luminosity, and have strong harmonic
structure. With {\it RXTE} observations the atoll LMXB have also been seen
to have these signals, although often the coherence is less and there
are other signals (See \cite{Wijnands00}). These QPO tend to be near 
in frequency to the break
frequency of band-limited white noise at low frequencies.

\begin{figure}[t]

\centerline{\epsfig{file=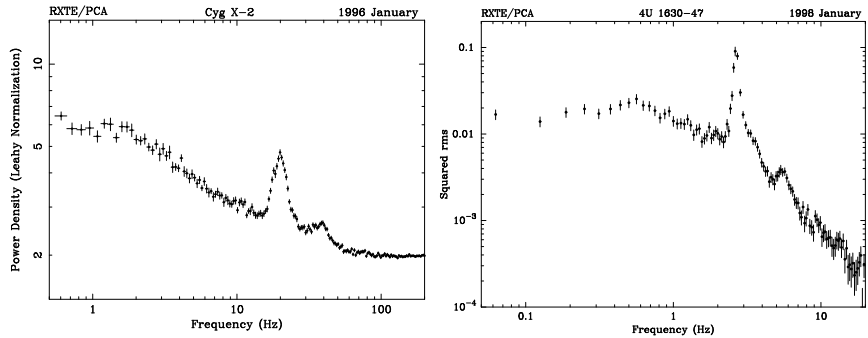,width=4.5in}}
\caption{Low Frequency Oscillations. (left) Horizontal Branch Oscillations
in Cygnus X-2.(right) Decahertz Oscillations in the Black Hole 
Candidate 4U 1630-47. These are cases with similar harmonic
structure, but a different relation to the break frequency of the lower 
frequency noise. States occur with very different harmonic structure.}

\label{fig5}

\end{figure}

The black hole transients had already exhibited very similar features
in Nova Muscae and GX 339--4 in the range 1-15 Hz.
They have very similar properties to the HBO. {\it RXTE} PCA observations
have found these QPO in the power spectra of most black hole
candidates \cite{Sw01}. 
Different origins have been discussed for the neutron star and black hole QPOs,
but their similarity is noted. Figure 5 
shows examples from a Z source and a black hole candidate 
(See \cite{Focke96,Dieters00}). 

The HBO were originally ascribed to a magnetic beat frequency model,
assuming the Kepler frequency and the spin were both not seen. 
Stella and Vietri identified them with the Lense-Thirring
precession (See \cite{Stella99}). 
They appear to have the correct quadratic relation to the
high frequency kilohertz QPO. But the magnitude was too high, by even a
factor of about four. Assigning them to twice the nodal frequency,
a reasonable possibility for the x-ray modulation, relieves the
problem in some cases, but still leaves a factor of two in many.
Psaltis argues that a  magnitude discrepancy of a factor of
two can be accommodated in situations where there is
actually complex hydrodynamic flow rather than single particle orbits
\cite{Psaltis01}.

In the case of the black holes, the energy spectra seem to distinguish
contributions of an optically thick disk and non-thermal, that is
``power-law'' emission, attributed to scattering of low energy photons
off more energetic electrons. This division of components is not
observationally so clear in the neutron star LMXB (There are many
plausible reasons for this: lower central mass and smaller inner disk,
X-rays generated on infall to the surface, possible spinning magnetic
dipole.) For the black hole transients, this low frequency QPO is
clearly a modulation of the power-law photons. However, there appear
to be a variety of correlations with the disk behavior, so that the
two components are clearly coupled.

In the case of the neutron stars Psaltis, Belloni, and van der Klis
\cite{PBvdK99} have noted that the HBO and the lower kilohertz
oscillation are correlated over a broad range of frequency (1--1000
Hz). Wijnands and van der Klis \cite{WvdK99} showed that {\em both}
the noise break and the low frequency QPO are correlated in the same
way for certain neutron stars and black hole candidates. Psaltis {\it
et al.} went on to point out that if some broad peaks in the power
spectra of some black holes were taken to correspond to the lower
kilohertz frequency in the neutron star sources, these points also
fell approximately on the same relation.

While the degree to which this relation was meaningful, given the
scatter in the points, selection effects, and distinctions of more than
one branch of behavior, recent work is suggestive  that in some way three
characteristic frequencies of the disk in a strong gravitational field
are significant, where these correspond to Kepler motion, precession
of the perihelion and nodal precession. There remain difficulties however
with specific assignments. 

It has often been noted that different interpretations implied weaker
features in the spectrum, for example modulation of frequencies by the
Lense-Thirring precession \cite{MarkovicL00} or excitation of higher
modes in the case of g-modes \cite{Wagoner98}.  
In the case of the neutron stars,
adding together large amounts of data to build up the statistical
signal, while Sco X-1 did not show sidebands \cite{Mendez00}, Jonker
et al. \cite{Jonker00} found evidence of sidebands at about 60 Hz to the lower
kilohertz frequency in three sources. The frequency separation is not
the same as the low frequency QPO in those sources although it is in
the same range and Psaltis argues is close enough that second order
effects can be responsible for the difference. It is not clear yet
whether the sidebands imply a modulation of the amplitude or whether
they represent a beat phenomenon and are one-sided.

\section*{Conclusions}

While it has not yet been possible to fit all the properties of LMXB
neatly into a model, it is hard to imagine alternatives for some
important results.  One of these is that in accordance with the theory
of General Relativity, there is an innermost stable orbit, such that
quasistatic disk flow does not persist inside it. Nuclear matter at
high densities does not meet such a stiff equation of state that the
neutron star extends beyond the ISCO. Instead the results suggest the
neutron star lies inside the ISCO for its mass.

The accretion flows for both neutron stars and black holes have
resonances which, from the observations, are apparently successfully
coupled to X-ray flux. QPO are observed with high coherence. They can
already be compared to assignments of various frequencies, but they do
not match exactly with the identifications that have been
made. However before it is possible to use it as diagnostic of
gravity, it is necessary to sort out further the physics of the
situations. 

Extending the measurements to signals an order of magnitude fainter
taxes even the abilities of {\it RXTE}. Continued observations are
pushing the limits lower by reducing statistical errors, but must deal
with intrinsic source variability on longer time-scales. Observations
are also being sought of especially diagnostic combinations of flux 
and other properties.

\end{document}